\documentclass[pra,notitlepage,showpacs,amsmath,amssymb,floatfix]{revtex4-1}

\usepackage{braket}
\usepackage{graphicx}
\usepackage{dcolumn}
\usepackage{bm}
\usepackage{textcomp}
\usepackage{upgreek}
\usepackage{lipsum}
\usepackage[colorlinks=true,citecolor=blue,linkcolor=blue]{hyperref}

\begin{document}
\title{An Underappreciated Radiation Hazard from High Voltage Electrodes in Vacuum}

\author{A. D. West}
\affiliation{Department of Physics, Yale University, 217 Prospect Street, New Haven, Connecticut 06511, USA}
\author{Z. Lasner}
\affiliation{Department of Physics, Yale University, 217 Prospect Street, New Haven, Connecticut 06511, USA}
\author{D. DeMille}
\affiliation{Department of Physics, Yale University, 217 Prospect Street, New Haven, Connecticut 06511, USA}
\author{E. P. West}
\affiliation{Department of Physics, Harvard University, Cambridge, MA 02138, USA}
\author{C. D. Panda}
\affiliation{Department of Physics, Harvard University, Cambridge, MA 02138, USA}
\author{J. M. Doyle}
\affiliation{Department of Physics, Harvard University, Cambridge, MA 02138, USA}
\author{G. Gabrielse}
\affiliation{Department of Physics, Harvard University, Cambridge, MA 02138, USA}
\author{A. Kryskow}
\affiliation{Environmental Health \& Safety, Harvard University, 46 Blackstone Street, Cambridge, Massachusetts 02139, USA}
\author{C. Mitchell}
\affiliation{Environmental Health \& Safety, Harvard University, 46 Blackstone Street, Cambridge, Massachusetts 02139, USA}
\date{\today}

\begin{abstract}
The use of high voltage (HV) electrodes in vacuum is commonplace in physics laboratories. In such systems, it has long been known that electron emission from an HV cathode can lead to bremsstrahlung X-rays; indeed, this is the basic principle behind the operation of standard X-ray sources. However, in laboratory setups where X-ray production is not the goal and no electron source is deliberately introduced, field-emitted electrons accelerated by HV can produce X-rays as an unintended hazardous byproduct. Both the level of hazard and the safe operating regimes for HV vacuum electrode systems are not widely appreciated, at least in university laboratories. A reinforced awareness of the radiation hazards associated with vacuum HV setups would be beneficial. We present a case study of a HV vacuum electrode device operated in a university atomic physics laboratory.  We describe the characterisation of the observed X-ray radiation, its relation to the observed leakage current in the device, the steps taken to contain and mitigate the radiation hazard, and suggest safety guidelines.

\end{abstract}
\newpage
\maketitle

\section{\label{sec:introduction}Introduction.}
High voltage (HV) devices are commonplace in physics laboratories. When used in air they do not present a radiation safety hazard, since the mean free path of any emitted electrons is always very short (${\sim}100$~nm) compared to the electrode gap, the accumulated kinetic energy before a collision is too small to produce ionizing radiation. HV electrodes in vacuum, however, can produce dangerous levels of radiation. Electrons emitted from a cathode surface can be accelerated across a HV vacuum gap, then rapidly stopped when they strike the opposing anode. The bremsstrahlung X-rays arising from this process were discovered over a century ago, and since then have been both thoroughly studied and widely used \cite{BremsstrahlungReview}. Standard X-ray tubes work on this very principle, and deliver high X-ray flux by deliberately accelerating electrons across the HV vacuum gap.

However, in many physics experiments, HV vacuum electrodes are used for entirely different type of purpose with no intent to make X-rays. For example, in the atomic, molecular and optical (AMO) physics community, the tasks of steering, focusing, slowing, and trapping atomic and molecular beams have been achieved using HV electrodes since the 1950's, up until the present \cite{Bennewitz1955,Ramsey1985,Meerakker2012,Bethlem1999,Maddi1999,Bethlem2002}. In these (and similar) types of experiments, the production of X-rays is entirely undesirable. Nevertheless, it is typical that some small, unwanted current of electrons continues to flow between the HV electrodes even when the apparatus is designed to minimize this effect; hence, unwanted X-rays will almost inevitably be produced at some level with any HV vacuum electrode setup.  

The potential hazard due to incidental X-ray production in HV vacuum electrode devices is often addressed in the radiation safety programs of laboratories and research institutions (e.g. \cite{SantaCruzXRay,WKUXRay,DOERadiologicalWorkerTraining}), and many researchers and technicians at nuclear facilities or particle accelerators are well aware of the hazard (e.g. \cite{IAEAXray,FRCM}). However, our experience, together with circumstantial evidence, suggests that awareness of the danger of X-rays arising from apparatus using HV vacuum electrodes may be less fully appreciated in university laboratories. In this paper, we wish to draw attention to the fact that essentially \textit{any} HV apparatus with in-vacuum electrodes can constitute an X-ray safety hazard, and that the associated dangers may not be appreciated by typical researchers in the field (particularly at universities). We begin with a discussion of the underlying mechanism leading to X-ray production, and give order-of-magnitude estimates for potential exposure while using equipment of this type; we then go on to discuss observations from a specific setup as a case study.

%

\section{\label{sec:oom_estimate}Basic Mechanisms and Order of Magnitude Dose Rate Estimate.}
In experiments using HV vacuum electrodes where X-ray production is \textit{not} a goal, it is typically preferred to minimize the current flowing between the HV electrodes.  In general, this so-called ``leakage current'' can take two physically distinct forms \cite{Bajic1989,Latham1995}.  In the first, current flows along the insulating spacers used to separate electrodes from each other and/or from ground.  With well-chosen materials, leakage through the bulk of the insulator is typically negligible.  However, current flowing on insulator surfaces is typically much larger. Suppressing such surface leakage currents requires careful consideration of surface cleanliness and geometrical layout. While frequently undesirable for a given experiment, these surface currents do not generally lead to significant X-ray production, since the electrons are continuously scattered on the surface and do not reach the anode with sufficient kinetic energy to produce bremsstrahlung.  

The second source of leakage current is the direct emission of electrodes from the cathode, which then accelerate through vacuum till they hit a surface at much lower electric potential (either an anode, or a grounded surface).  While this process is frequently referred to as ``field emission'' --- a phrase which refers to a specific physical mechanism for electron emission --- in fact there are a variety of physical mechanisms that that can lead to leakage currents flowing through vacuum \cite{Bajic1989,Latham1995}.  We refer to them collectively as ``vacuum emission''.   Leakage currents from vacuum emission are specifically responsible for the bremsstrahlung X-rays that are the subject of this paper.  

The magnitude of vacuum emission currents in HV devices can vary widely, and can depend strongly on the history of the electrodes' use. It is widely known that some methods of ``conditioning'' vacuum HV electrodes can lead to extremely small vacuum emission currents (in the pA range or lower even for applied voltages of many tens of kV) \cite{Bajic1989,Latham1995}.  However, it appears typical that unconditioned electrodes can have can have vacuum emission currents of 1~$\upmu$A or more \cite{Williams1972,BastaniNejad2014,Sudarshan1988,HudsonThesis} and indeed, this is typical of the levels observed in the apparatus described here.  As we explain below, this range of current can easily be sufficient to produce a significant radiation hazard.  We note in passing that, in any given HV setup, it is typically easy to measure total current but far less clear how to distinguish (radiation-benign) surface currents from (radiation hazard-inducing) vacuum emission currents.  Out of an abundance of caution, researchers should assume that any measured HV current could be a pure vacuum emission current that leads to bremsstrahlung X-rays, unless there is clear evidence to the contrary. In our setup, after some early efforts to suppress surface currents, it appears that most of the typically 1--10 $\upmu$A currents were indeed vacuum emission currents.

With this in mind, we next estimate the dose rate due to a vacuum emission current of ${\sim}1~\upmu$A (${\sim}10^{13}~e^-$~s$^{-1}$) accelerated across a 60~kV potential difference (similar to the currents observed and voltages used in our setup) and then incident on an iron surface (similar to the stainless steel anodes in our setup).  The total power in the electron current is ${\sim}60$~mW. For 60~keV electrons impacting on iron, ${\sim}0.2~\%$ of the energy is converted into bremsstrahlung \cite{NISTestar}, giving ${\sim}120~\upmu$W of X-ray power. We assume a photon energy spectrum following Kramers' law \cite{Kramers1923}. In traversing the distance from the in-vacuum electrode to the position of a lab worker, we assume that the photons pass through 3~mm of aluminum (the vacuum chamber wall). Using standard tables of X-ray absorption coefficients \cite{NISTXray}, we find that ${\approx}20~\%$ of the photon energy passes out of the chamber, with a mean transmitted X-ray energy of ${\approx}40$~keV. The total power emitted from the chamber is then ${\sim}20~\upmu$W. Although bremsstrahlung emission is typically mildly directional \cite{Jackson} we assume isotropic emission for simplicity. As a conservative estimate we assume that a person located at the closest accessible position, ${\sim}10$~cm from the source of radiation, subtends a solid angle of ${\sim}2\pi$. The dose rate received is then ${\sim}400~\upmu$Sv~hr$^{-1}$\footnote{For convenience, note that 1~nW of radiation absorbed by a body of 100~kg is equivalent to a dose rate of 36~nSv~hr$^{-1}$ (assuming a quality factor and tissue weighting factor of 1).}. In summary, we estimate the dose rate per current at a distance of 10~cm from a point of bremsstrahlung production, $D_I$, to be
\begin{equation}
D_I{\sim}400~\upmu{\rm Sv~hr}^{-1}~\upmu{\rm A}^{-1}.
\end{equation}
For comparison, the annual dose limit for radiation workers (members of the public) corresponds to a dose rate of 25~$\upmu$Sv hr$^{-1}$ (500~nSv hr$^{-1}$) assuming 2,000 hours (approximately one work year) of exposure. 

This estimate makes it possible to provide rough guidelines for safe operating conditions in an apparatus with vacuum HV electrodes.  For example, when working with 60 kV potential differences as in the estimate above, vacuum emission currents as low as 50 nA can lead to a significant hazard for long-term exposure.  Repeating this calculation at other voltage levels suggests that currents near or above $1~\upmu$A, when used with voltage in the 40~kV range, also represent a significant potential hazard. By contrast, at a lower voltage of $\sim$20~kV and for current of $1~\upmu$A, the dose rate is estimated to be less than the public limit quoted above. Of course, we emphasize that all such estimates should be interpreted conservatively, and that direct measurement of dose rates is always the most accurate method in assessing potential radiation hazards.

With this discussion in mind, we now go on to describe a particular experience with X-ray production from an apparatus with HV vacuum electrodes in our laboratory, as a case study.

\section{Case Study: Electrostatic Lens}
\subsection{Construction and operation}
\begin{figure}[!ht]
\centering
\includegraphics[width=0.7\linewidth]{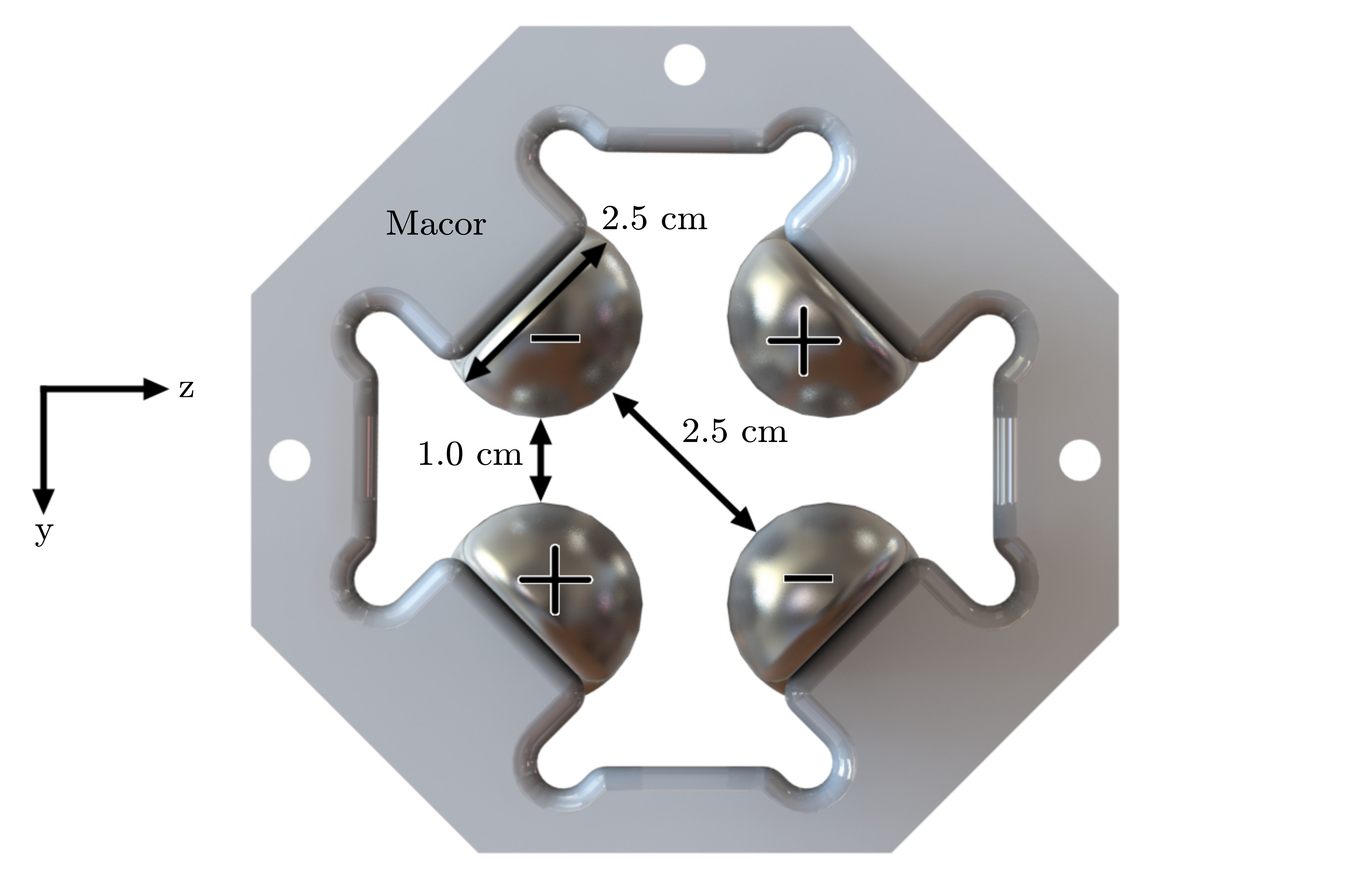}
\caption{Simulated end-on (oriented along $\hat{x}$) view of the HV electrodes and one of the macor ceramic mounts which support them. The $y$-axis points downwards in the laboratory. Plus and minus signs indicate the polarity of the applied HV.}
\label{fig:electrodes_and_macor}
\end{figure}

The apparatus we studied is a quadrupole electrostatic lens designed to focus a beam of $^{232}$ThO molecules as part of the Advanced Cold Molecule Electron EDM (ACME) experiment to measure the electric dipole moment of the electron \cite{Baron2014}. It consists of a set of four stainless steel (316L) electrodes, of diameter 2.5~cm and length 60~cm. Electrodes of opposite voltage polarity are separated by a minimum distance of ${\approx}1$~cm, as shown in Fig.~\ref{fig:electrodes_and_macor}, and are fastened to two macor ceramic mounts via polyether ether ketone (PEEK) plastic screws.

The electrodes were prepared as follows: metal stock was annealed, cold drawn and then ground to the appropriate diameter, after which it was cut to length and rounded to a spherical profile at either end. Some small features were machined to facilitate supporting the electrodes and making the electrical connections. Following machining, the electrodes were electropolished and then cleaned in an ultrasonic bath sequentially with Alconox, acetone, isopropanol, and deionized water. Any residual dust or debris was removed during assembly with optics tissue or an ultra pure gas duster.

Electrical connections to the electrodes are made via separate ellipsoidal stainless steel connectors (diameter ${\approx}1.9$~cm, length ${\approx}2.3$~cm) which attach via set screws directly onto flat surfaces of the electrodes on one side, and clamp onto a Kapton-coated wire via a set screw on the other side. Similar connectors are also used to connect the other ends of the wires to HV vacuum feedthrough pins.

The HV is supplied by two DC supplies which are continuously variable from 0 to $\pm30$~kV and can provide up to 250~$\upmu$A of current. Each supply provides the voltage for two (diagonally opposite) electrodes of the same polarity. The apparatus was designed to operate with voltages of $\pm30$~kV applied. Unless otherwise specified, all tests were performed with this voltage. A vacuum of less than 1~$\upmu$Torr is maintained in the chamber by a turbomolecular pump.

\subsection{Investigation of radiation emitted}
\subsubsection{Discovery}
X-ray radiation from the apparatus was first discovered incidentally during an independent radiation study in the same laboratory, in which activity from the $^{232}$Th isotope was being measured with a GM probe and survey meter. High count rates (${\sim}10^6$~cpm) were observed when surveying near the HV apparatus while it was powered on, with a measured leakage current of ${\sim}1$--10~$\upmu$A. At this point, the apparatus was moved to a separate lab where suitable shielding (lead blankets), radiation monitoring and personnel controls could be put in place. Shortly following the discovery, the Department of Environmental Health and Safety issued a warning to all on-campus radiation users and to any potential high voltage users, reiterating the possible hazards of working with high voltages in a vacuum environment. The case was presented to the Radiation Safety Committee and additional detail on this type of hazard was added to the relevant safety training programs.

After initial surveying, measurement of the radiation emitted was performed primarily using a GM-probe based dosimeter. A number of experimental parameters were monitored simultaneously, including the voltage and current provided by the power supplies, and the pressure inside the vacuum chamber. The remainder of this section describes the measurements taken.

\subsubsection{Initial Surveying}
\label{sec:initial_surveying}
\begin{figure*}[!ht]
\centering
\includegraphics[width=0.7\linewidth]{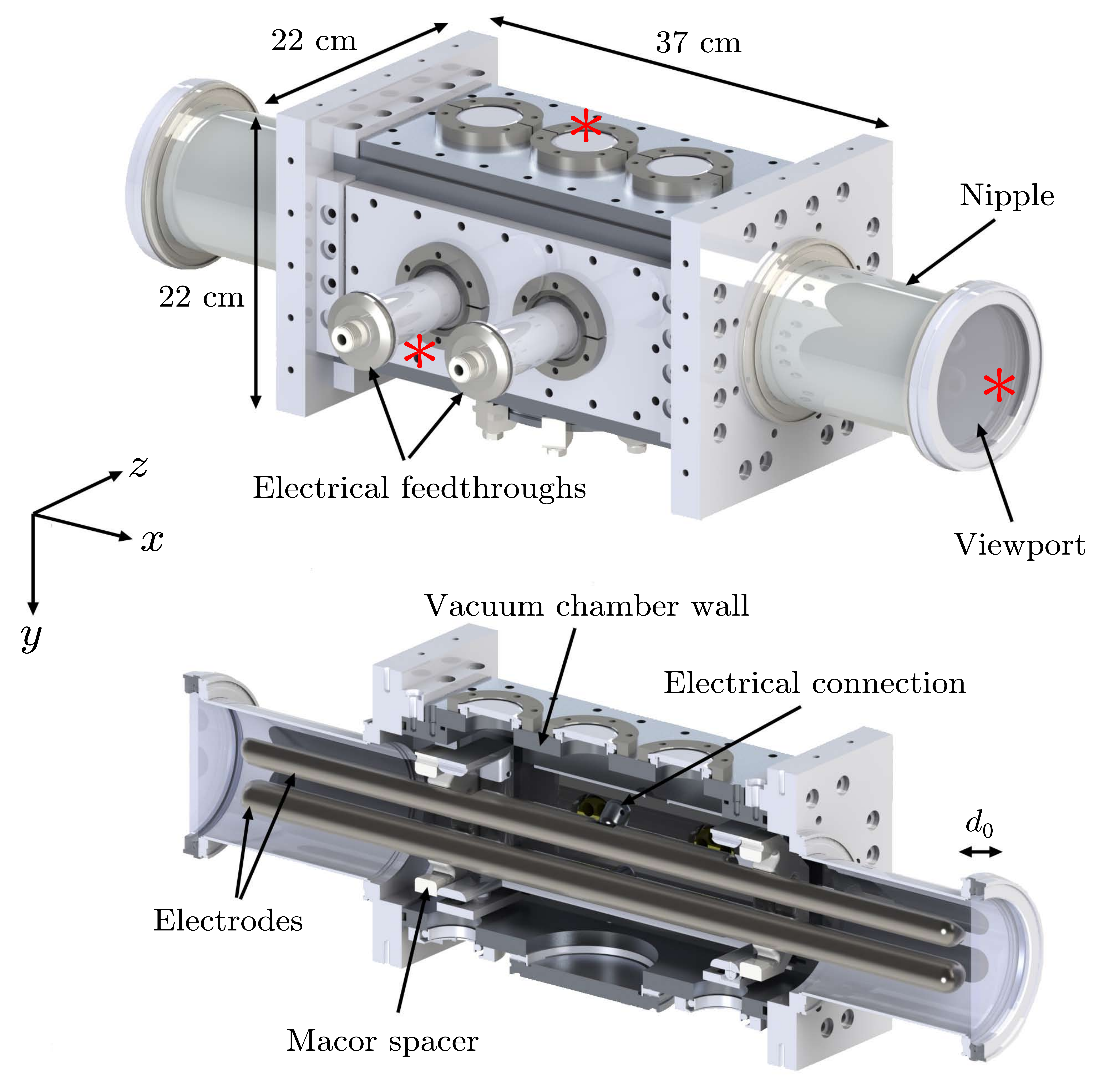}
\caption{A schematic of the experimental apparatus, showing the vacuum chamber containing the electrodes. The $x$-axis is aligned with the electrodes, while the $y$-axis points downwards. The bottom image is a cutaway in the $xy$ plane such that only two of the four electrodes are shown. The electrodes are oriented with their length along the $x$-axis, centered on $x=0$, and extend around 10~cm into the nipples on either end. The center of the quadrupole configuration shown in Fig.~\ref{fig:electrodes_and_macor} is positioned at $y=z=0$, i.e. centered within the chamber/nipples. Red asterisks indicate positions where particularly high count rates were observed during initial surveying. $d_0$ labels the estimated distance of a source of radiation, as described in Section~\ref{sec:dosimetry} (see Figure~\ref{fig:dose_rate_vs_distance}).}
\label{fig:apparatus_schematic}
\end{figure*}

Measurements were carried out with a GM-probe (Ludlum 44-9) to localize the source of radiation and provide initial estimates of the dose rate. During the measurements, leakage currents of ${\sim}1$--10~$\upmu$A were observed. Three locations with particularly high count rates were observed. These positions and the overall apparatus are shown in Fig.~\ref{fig:apparatus_schematic}. 
We observed variation in the count rate by a factor of several hundred while moving around the apparatus. This variation is significantly larger than expected due to any variation in distance from the electrodes or in thickness of the vacuum chamber walls.

Using the count rate, $\dot{n}\approx2$~Mcpm, measured with the GM probe, it was possible to make an initial estimate of the dose rate. We calculated the expected energy spectrum both with commercial software \cite{Wu2009} and custom code; we describe here the procedure using the latter. As in our simple estimates above, we assumed a photon number energy spectrum, $N(E)$, following Kramers' law \cite{Kramers1923} for the case of 60~keV electrons, modified by the energy-dependent transmission by the vacuum chamber walls (aluminum), $T(E)$ \cite{NISTXray}. The dose rate, $\dot{D}$, can then be shown to be given by
\begin{equation}
\dot{D}=\frac{\dot{n}}{S}\frac{\int N(E)T(E)\mu(E)E\enspace dE}{\int N(E)T(E)R(E)\mu(E)E\enspace dE},
\end{equation}
where $S$ is the GM probe's sensitivity to Cs-137 gamma rays (count rate per dose rate) \cite{Ludlum449} and $R(E)$ is the sensitivity of the probe at an energy $E$, relative to the sensitivity to Cs-137 gamma rays \cite{Ludlum449}, and $\mu(E)$ is the mass attenuation coefficient for air at an energy $E$. From this we estimate a dose rate of ${\approx}3$~mSv hr$^{-1}$, which is in agreement with the order of magnitude estimate provided in Section~\ref{sec:oom_estimate} for a leakage current due to field emission of ${\approx}7~\upmu$A. In this calculation we assumed that the radiation is isotropic, meaning the quoted dose rate is likely an overestimate given the observed directionality.

\subsubsection{Dosimetry Surveying}
\label{sec:dosimetry}
More detailed quantitative measurement of the spatial variation in the measured X-ray flux was performed in two ways. The first method used dosimeter badges (Landauer Luxel+) which were placed around the device at several locations while the apparatus was operated continuously for approximately 40~mins. During this time, the leakage current was again ${\sim}1$--$10~\upmu$A. The positions of the badges and the resulting doses and dose rates are shown in Table~\ref{tab:badge_positions} (badge analysis was performed assuming radiation solely from X-rays). We define a coordinate system with its origin at the center of mass of the four electrodes. The $x$-axis is aligned with the long dimension of the electrodes and the $y$-axis is downward (see Figures~\ref{fig:electrodes_and_macor} and \ref{fig:apparatus_schematic}).
\begin{table*}[!ht]
\centering
\begin{tabular}{ccccccc}
\textbf{Badge} & \textbf{Position (cm)} & \textbf{DDE ($\upmu$Sv)} & \textbf{LDE ($\upmu$Sv)} & \textbf{SDE ($\upmu$Sv)} & \textbf{DDRE ($\upmu$Sv hr$^{-1}$)}\\\hline
1 & (46, -10, 3) & 10 & 10 & 0 & 20\\ 
2 & (46, 0, 23) & 10 & 10 & 10 & 20\\ 
3 & (51, 10, -20) & 0 & 0 & 0 & 0\\ 
4 & (0, -13, -3) & 1700 & 1890 & 1960 & 2550\\ 
5 & (0, -10, 20) & 180 & 180 & 180 & 270\\ 
6 & (10, 0, -25) & 260 & 260 & 240 & 390\\ 
7 & (-56, 10, -3) & 0 & 0 & 0 & 0\\ 
8 & (104, 0, 15) & 30 & 30 & 20 & 50\\ 
9 & (86, -8, 69) & 0 & 0 & 0 & 0\\ 
\end{tabular}
\caption{Positions surveyed with badge dosimeters. `DDE', `LDE' and `SDE' refer to deep dose equivalent, lens dose equivalent and shallow dose equivalent, respectively. `DDRE' is the deep dose rate equivalent. Doses listed as zero were below the minimal measurable dose of 10~$\upmu$Sv. Leakage currents of ${\sim}1$--$10~\upmu$A were present during these measurements.}
\label{tab:badge_positions}
\end{table*}

Out of the positions surveyed with badge dosimeters, three revealed particularly high dose rates. These were approximately located at coordinates $(0,-13,-3)$~cm, $(0,-10,20)$~cm, and $(10,0,-25)$~cm, all of which refer to positions near the center of the apparatus. Surprisingly, a large dose rate was \textit{not} observed by badges at one of the areas of highest count rate found during initial surveying, located near the viewport at one end of the electrodes ($x\approx+40$~cm, right-most asterisk in Fig.~\ref{fig:apparatus_schematic}). Further measurements showed that this was because the X-ray flux in that region was highly directional, being well-aligned with the $x$-axis. (For example, at an $x$ position corresponding to directly outside one of the viewports, the dose rates at $y\approx0$~cm and $y\approx15$~cm differed by a factor of ${\sim}10$.) The shielding provided by the electrodes themselves may have contributed to this directionality.

The second method for quantitative study of the dose rate utilized a handheld dosimeter (RadEye B20-ER). These measurements supported the conclusion from initial surveying and dosimeter badge readings that the source of radiation was well-localized at two positions, one near one of the viewports on the $x$-axis and the other near the center of the vacuum chamber where $x=0$ (cf. Figure~\ref{fig:apparatus_schematic}). After careful surveying, a maximum dose rate on contact with the vacuum chamber of ${\approx}2.5$~mSv hr$^{-1}$ was measured, although this maximum was observed to vary over the range ${\sim}0.1$--2.5~mSv hr$^{-1}$ on timescales of hours to days. During this time the leakage current was ${\sim}1$--10~$\upmu$A. We note that the maximum dose rate from the handheld dosimeter also agrees with our other estimates of dose rate.

\begin{figure}[!ht]
\centering
\includegraphics[width=0.75\linewidth]{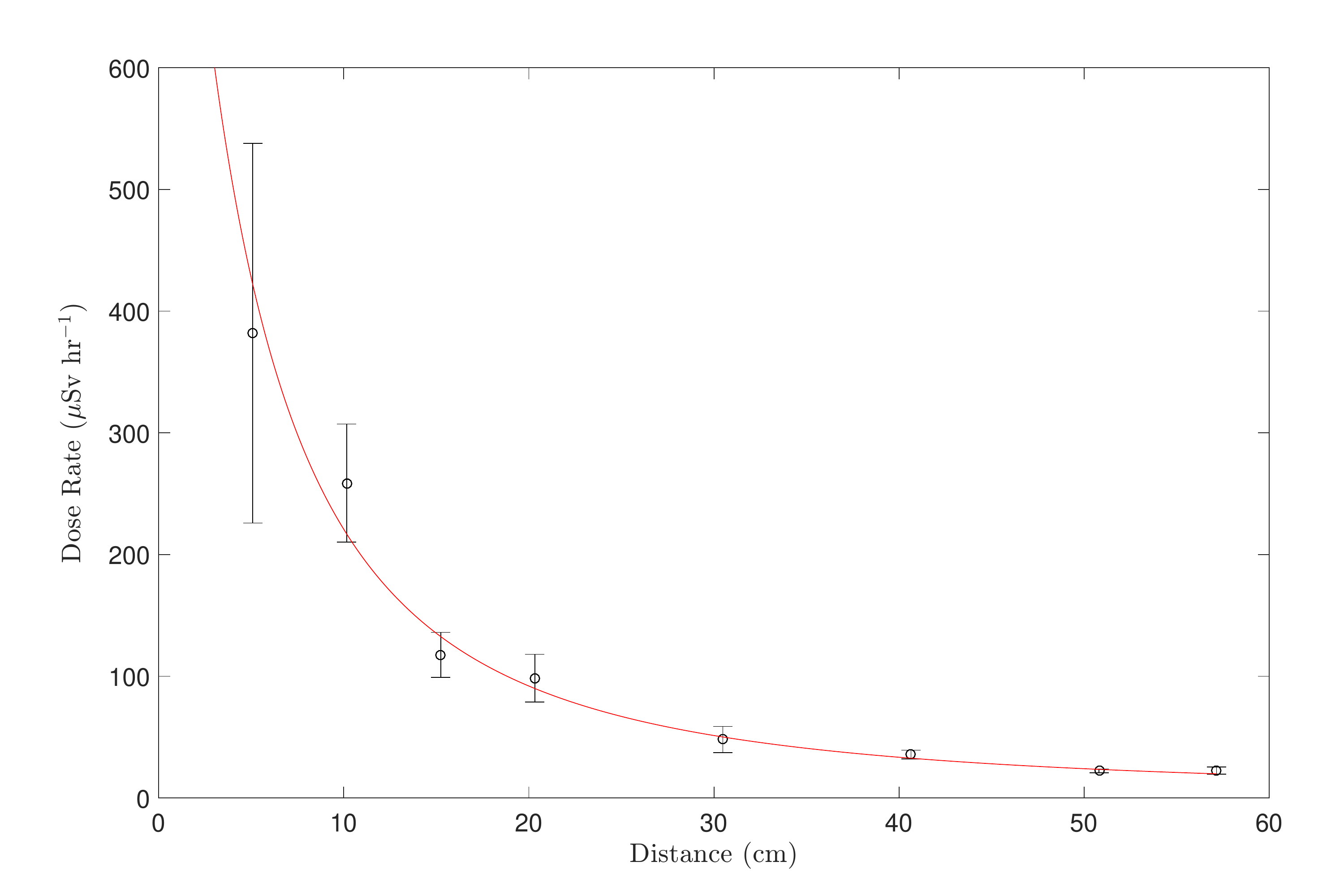}
\caption{Measured dose rate $D$ as a function of distance $d$ along the $+\hat{x}$ axis. Zero is located at the viewport surface. Black data points represent measurements and the red line is the result of a fit to a curve of the form $D(d)=D_0/(d-d_0)^2$, where $D_0$ is an adjustable fit parameter and the best fit gives $d_0\approx-8$~cm. Due to drifts in the count rate, fluctuations in the data are not anticipated to be Gaussian, so the error bars represent the overall range of measurements at a particular distance. During these data, the leakage current was ${\sim}1$--10~$\upmu$A.}
\label{fig:dose_rate_vs_distance}
\end{figure}

In order to localize the source of the radiation production more precisely, measurements were carried out around the regions of highest observed dose rate. Directions along which the radiation was peaked were identified and the dose rate was then measured as a function of distance. Example data are shown in Fig.~\ref{fig:dose_rate_vs_distance}.
Given observations of the localized character of the radiation, we chose to model the radiation source as a point. 
As shown in Fig.~\ref{fig:dose_rate_vs_distance}, this model fits the data well and suggests that there is a source of radiation located 8~cm inside the viewport. This position is physically plausible, since it is within the length of the electrodes (roughly 2~cm from their end). Similar analysis indicated another localized source of radiation close to $x=0$, the center of the electrodes' lengths and where the electrical connections are made, with comparable measured dose rates. This suggests that close examination of these connections may help mitigate electron emission.



\subsubsection{X-ray Energy Spectrum}
To verify our understanding of the X-ray emission process, and to predict what shielding methods would be sufficient to ensure safe use of the device, we performed a qualitative characterization of the X-ray spectrum. In the absence of an X-ray spectrometer, this was achieved by examining the variation in the count rate, as measured by the handheld dosimeter, as a function of the shielding present. Sheets of ${\approx}0.8$~mm thick aluminum were stacked to give thicknesses of up to ${\approx}1.4$~cm and placed in front of the dosimeter, which was located around 10~cm from the end of the vacuum chamber, in the $+\hat{x}$-direction. For each thickness the count rate was recorded 30 times over a period of ${\sim}2$~mins. The thickness of the aluminum was varied in a random order. The leakage current (typically ${\sim}10~\upmu$A for these data) was measured simultaneously.

\begin{figure}[!ht]
\centering
\includegraphics[width=0.75\linewidth]{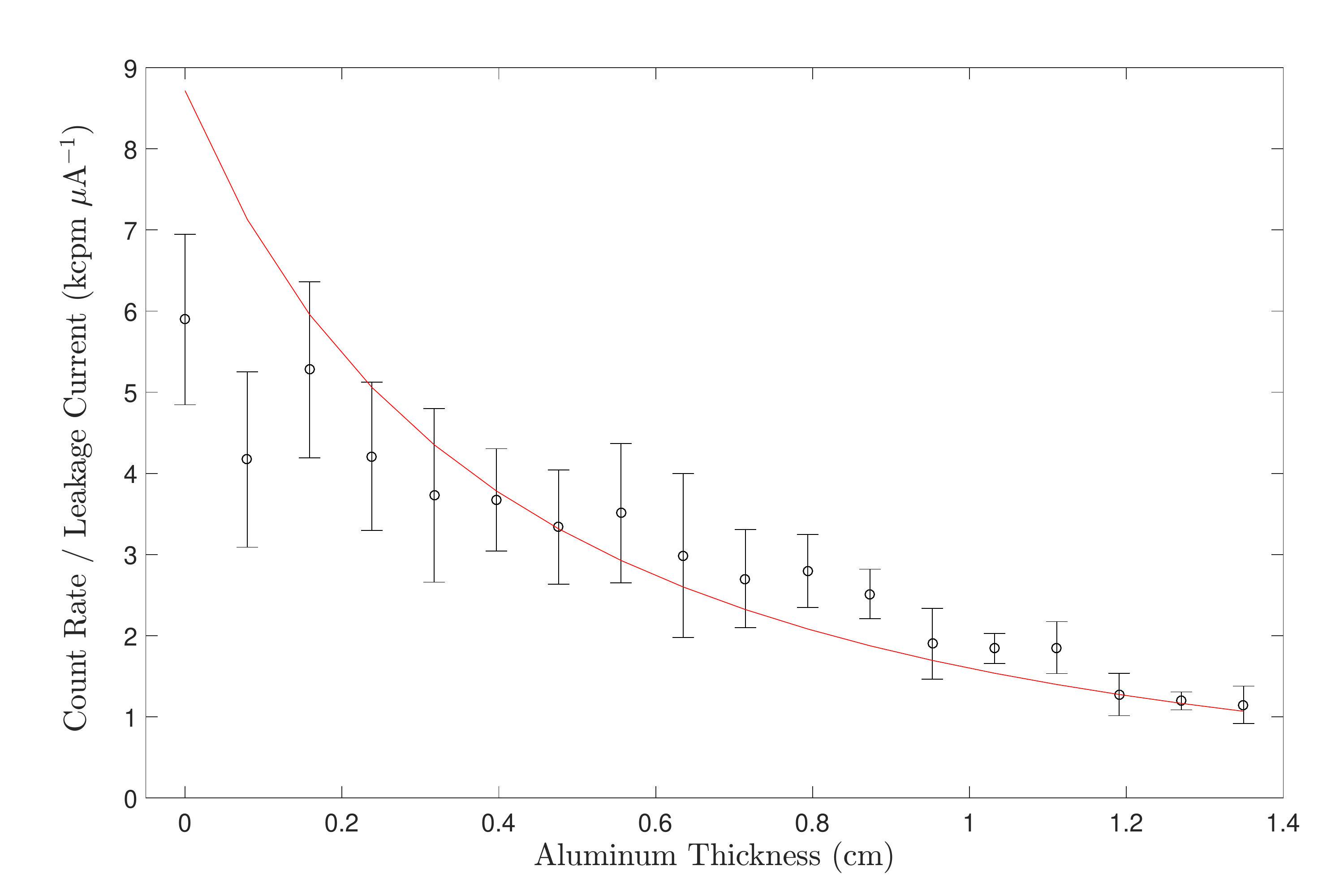}
\caption{Observed dose rate (normalized to leakage current) as a function of thickness of aluminum shielding. Black data points represent measurements and the red line is the result of calculation assuming a 60~keV electron energy, fitting the overall scaling as a free parameter. Due to drifts in the count rate, fluctuations in the data are not anticipated to be Gaussian, so the error bars represent the overall range of measurements at a particular thickness.}
\label{fig:countrate_vs_al_thickness}
\end{figure}

Measured data were then compared to a simulated spectrum. The photon energy distribution was calculated as described in Section~\ref{sec:initial_surveying}, accounting for absorption by the vacuum chamber viewport and aluminum plates. To convert from photon number to measured count rate requires knowledge of the energy dependence of the dosimeter response, which is approximately constant \cite{RadeyeManual}. This simulation was then repeated while varying the thickness of aluminum shielding. A least squares fit of the resulting curve to the data was performed while varying an overall scaling factor. The results are shown in Fig.~\ref{fig:countrate_vs_al_thickness}.

The qualitative dependence of the dose rate on shielding thickness is fairly well reproduced under the assumption of a bremsstrahlung spectrum, crudely validating our understanding of the spectrum of X-rays emitted from our apparatus. The discrepancy between data and simulation is likely due to inaccuracy in our model --- for example there may be additional shielding of X-rays due to the electrodes themselves.

\subsubsection{Dependence on Voltage and Leakage Current}
The emission of bremsstrahlung radiation from our apparatus is a clear indication that electrons are flowing between the electrodes, though the apparatus was designed with the goal of producing no such leakage current. Minimizing this leakage current is an important step in mitigating the radiation hazard. 

\begin{figure}[!ht]
\centering
\includegraphics[width=0.8\linewidth]{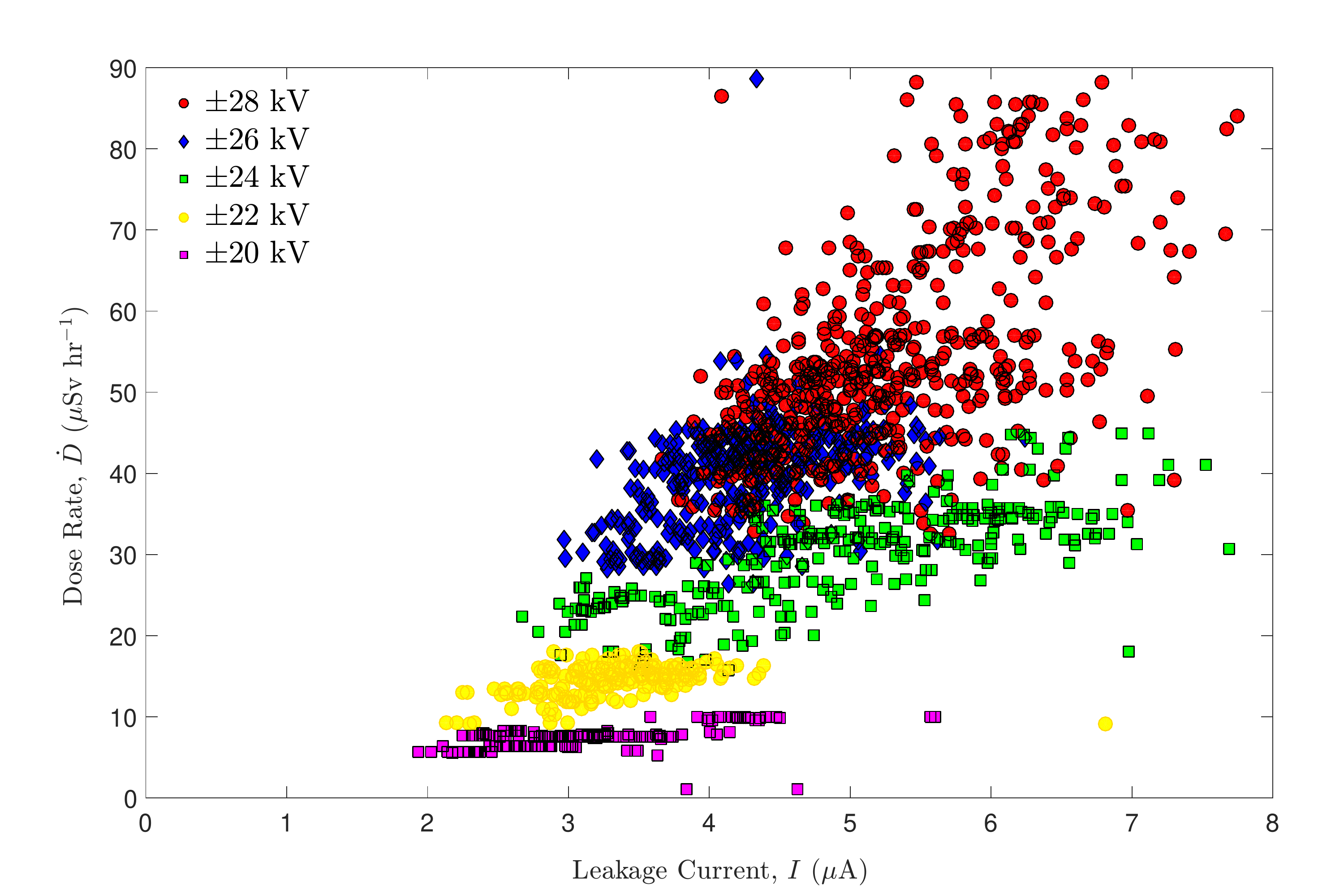}
\caption{Measured dose rate $\dot{D}$ as a function of the leakage current $I$ between the electrodes. Each data series, represented by different colors/symbols, corresponds to a different voltage applied to the electrodes. Measurement was performed at position $(x,y,z)=(45,0,0)$~cm, i.e. a distance of ${\approx}15$~cm from the end of the electrodes, in the $+\hat{x}$ direction.}
\label{fig:doserate_vs_current_voltage_combined}
\end{figure}

\begin{figure}[!ht]
\centering
\includegraphics[width=0.75\linewidth]{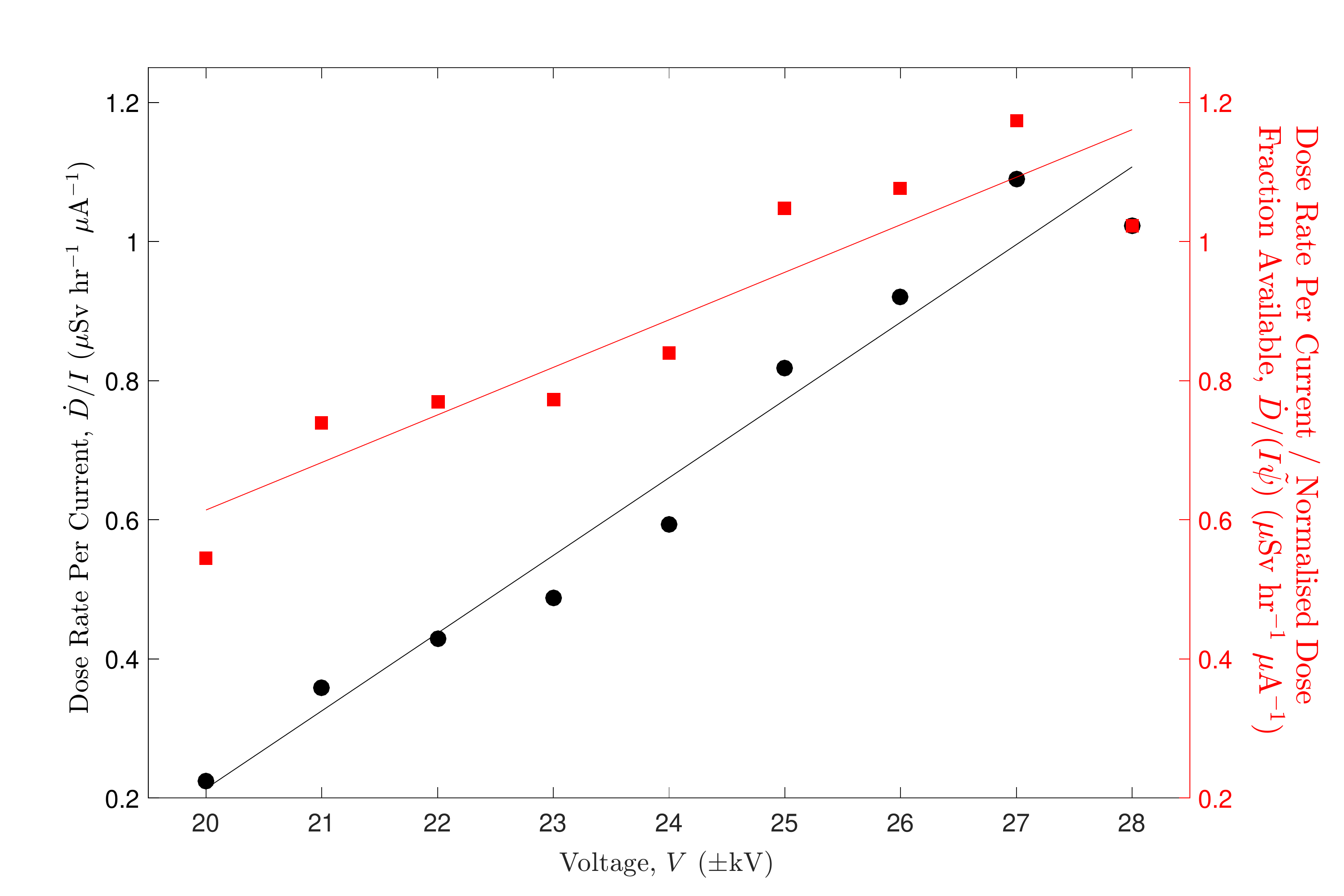}
\caption{Black circles: Measured dose rate divided by leakage current present, as a function of the applied voltage, produced by performing a least-squares fit to data such as that shown in Fig.~\ref{fig:doserate_vs_current_voltage_combined}. Red squares: black circle data points divided by a voltage-dependent factor which accounts for changes in the efficiency of bremsstrahlung production and detection --- see main text for description. The straight lines shown are fits to guide the eye.}
\label{fig:dose_rate_per_current_vs_voltage}
\end{figure}

To further characterize the radiation emission, we measured the dose rate from X-rays while varying the applied voltage and monitoring the current drawn from the power supplies. Again, the measurement was performed on the $x$-axis, ${\approx}10$~cm from the viewport at one end of the vacuum chamber. For applied voltages below $\pm20$~kV, the measured dose rate was consistent with background. The resulting data are shown in Fig.~\ref{fig:doserate_vs_current_voltage_combined}.
For each value of the applied voltage, the dose rate increases with the observed leakage current as expected. As the voltage is increased, there is also a clear increase in the dose rate per unit of leakage current. Performing a linear fit to determine the slope of the dose rate vs.\ current line for each voltage value allowed us to quantify this variation. The resulting data are shown as black squares in Fig.~\ref{fig:dose_rate_per_current_vs_voltage}. We see that there is a roughly fivefold increase in the measured dose rate per leakage current when the voltage increases from $\pm20$ to $\pm28$~kV.

To understand better the origin of this increase, we considered possible sources of variation in dose rate per leakage current. In general we can write the measured dose rate, $\dot{D}(V)$ as
\begin{equation}
\dot{D}(V)\propto2IV\times f\times C(2Ve)\times\frac{\int_0^{2Ve}N(E)ET(E)\mu(E)\enspace dE}{\int_0^{2Ve}N(E)EdE}\equiv fI\psi(V),
\label{eq:dosepropto}
\end{equation}
where $\pm V$ and $I$ are the measured electrode voltages and leakage current, $e$ is the electron charge, $C$ is the theoretical bremsstrahlung radiation yield \cite{NISTestar}, $N$ is the theoretical bremsstrahlung photon number distribution \cite{Kramers1923}, $\mu$ is the mass attenuation coefficient of air \cite{NISTXray}, $T$ is the calculated transmission of X-rays through the vacuum chamber viewport \cite{NISTXray} and $f$ is the fraction of electrical power from the leakage current that is associated with field-emitted electrons. Our goal is to deduce $f$ (up to a constant of proportionality). Equation~\ref{eq:dosepropto} can be understood intuitively as follows: $2IVf$ is the electrical power associated with field-emitted electrons, $C$ is the fraction of this energy that is converted into bremsstrahlung, and the term in the integral represents the fraction of the bremsstrahlung power which escapes the vacuum chamber and is recorded as a dose. In particular, both the radiation yield and the fraction of energy transmitted through the vacuum chamber increase with applied voltage. Using Equation~\ref{eq:dosepropto} we computed $\tilde{\psi}(V)\equiv\psi(V)/\psi(56~{\rm keV})$. Dividing the dose rate per current ($\dot{D}/I$) by $\tilde{\psi}(V)$ yields a quantity proportional to $f$. Plots of both $\dot{D}/I$ and $\dot{D}/(I\tilde{\psi})$ vs. applied voltage $V$ are shown in Figure~\ref{fig:dose_rate_per_current_vs_voltage}. As seen there, our data indicate that the fraction of the leakage current electrical power that is converted into field emission, $f$, increases with the applied voltage. This suggests that in our system field emission has a stronger dependence on voltage than surface current production does (surface current \cite{Gleichauf1951,Miller1989,Shannon1965,Kuffel1972,Grzybowski1983,Korzekwa1989} is a type of leakage current that does not produce bremsstrahlung).


\subsection{Reduction of Radiation}
The mitigation of leakage current and electrical breakdown in HV systems is an extensively studied problem \cite{Latham1995,Hawley1968,Faircloth2014}. It is known that field emission is enhanced by imperfections in the surface of the electrodes, either from impurities and defects \cite{Bommakanti1990} or from inherent roughness of the conducting surfaces \cite{Williams1972,Kato2007,Choulkov2005}. For these reasons it is important to ensure that all surfaces that are at HV are as smooth, uniform and clean as possible. Following such preparation, additional reduction of leakage current can be achieved through various methods of conditioning, such as spark conditioning \cite{Hatfield1988,Kuffel1972}, glow discharge conditioning \cite{Dylla1988,Anjali1994} or gas conditioning \cite{Yamamoto1989,BastaniNejad2014,Bajic1989}. The underlying principle behind all of these methods is the removal of surface imperfections that cause concentrations of electric field, usually microscopic metallic protrusions or foreign material, via destructive processes such as ion bombardment. It is of particular note that the use of emitted radiation has previously been recognized as an invaluable tool for diagnosing efficacy of conditioning \cite{Hernandez-Garcia2009}. 

In our case we observed that spark conditioning was useful in preventing surface flashover; however, the vacuum leakage current and associated radiation remain large enough to be of concern in our setup at the moment. Testing of the HV apparatus suggested that electron emission and bremsstrahlung were caused by the electrodes themselves, rather than the connectors, wires or feedthroughs. Future work will focus on finding a surface preparation method that will mitigate the problem.

\section{Conclusion}
We have observed and characterized the generation of bremsstrahlung X-rays from a device with HV vacuum electrodes, based in a university atomic physics laboratory. Although we observed no radiation at applied voltages below $\pm20$~kV, we observed a dose rate as large as 2.5~mSv hr$^{-1}$ immediately outside the apparatus when leakage currents of ${\sim}1$--10~$\upmu$A were present at applied voltage in the $\pm20$--$\pm30$~kV range. This dose rate was consistently measured via three different methods, and agrees with a simple estimate based on an assumption that the measured leakage current is entirely due to vacuum emission in the appratus. This level of radiation represents a substantial hazard under conditions frequently used in university physics laboratories. We observed the radiation to be highly directional and its sources to be well-localized. This makes it clear that single-point radiation monitoring around apparatus of this type may not be sufficient to understand the associated hazard.
HV devices operating in vacuum in a similar range of operating parameters are widely used in university physics laboratories.  Given an apparent unawareness of the associated radiation hazard, at least within parts of the community of university physics researchers, it is important to highlight this issue as a potentially unrecognized safety hazard.

\bibliography{library}

\begin{thebibliography}{10}

\bibitem{BremsstrahlungReview}
Edited by~C.~A.~Quarles and R.~H. Pratt.
\newblock Bremsstrahlung: Theory and experiment.
\newblock {\em Radiation Physics and Chemistry}, 75(10):1113--1430, 2006.

\bibitem{Bennewitz1955}
H.~G. Bennewitz, W.~Paul, and Ch. Schlier.
\newblock Fokussierung polarer molek\"{u}le.
\newblock {\em Z. Phys.}, 141:6--15, 1955.

\bibitem{Ramsey1985}
Norman Ramsey.
\newblock {\em {Molecular beams}}.
\newblock Oxford University Press, 1985.

\bibitem{Meerakker2012}
S.~Y.~T. van~de Meerakker, H.~L. Bethlem, N.~Vanhaecke, and G.~Meijer.
\newblock Manipulation and control of molecular beams.
\newblock {\em Chem. Rev.}, 112:4828--4878, 2012.

\bibitem{Bethlem1999}
H.~L. Bethlem, G.~Berden, and G.~Meijer.
\newblock Decelerating neutral dipolar molecules.
\newblock {\em Phys. Rev. Lett.}, 83:1558--1561, 1999.

\bibitem{Maddi1999}
J.~A. Maddi, T.~P. Dinneen, and H.~Gould.
\newblock Slowing and cooling molecules and neutral atoms by time-varying
  electric field gradients.
\newblock {\em Phys. Rev. A}, 60:3882--3891, 1999.

\bibitem{Bethlem2002}
H~L Bethlem, F~M~H Crompvoets, R~T Jongma, S~Y~T van~de Meerakker, and
  G~Meijer.
\newblock {Deceleration and trapping of ammonia using time-varying electric
  fields}.
\newblock {\em Physical Review A}, 65(5):053416, May 2002.

\bibitem{SantaCruzXRay}
Environmental Health and Safety.
\newblock Analytical {X-ray} safety workbook.
\newblock Technical report, University of California, Santa Cruz, 2001.

\bibitem{WKUXRay}
Radiation~Safety Office.
\newblock Radiation safety training.
\newblock Technical report, Western Kentucky University, 2015.

\bibitem{DOERadiologicalWorkerTraining}
U.S.~Department of~Energy.
\newblock {DOE} handbook --- radiological worker training.
\newblock Technical report, 2008.

\bibitem{IAEAXray}
International Atomic~Energy Agency.
\newblock Radiation safety of gamma, electron and {X-ray} irradiation
  facilities.
\newblock Technical report, 2010.

\bibitem{FRCM}
Health Environmental~Safety and Quality.
\newblock Fermilab radiological control manual.
\newblock Technical report, Fermilab, 2015.

\bibitem{Bajic1989}
S.~Bajic, A.~M. Abbot, and R.~V. Latham.
\newblock The influence of gap voltage temperature and gas species on the gas
  conditioning of hv electrodes.
\newblock {\em IEEE Trans. Electr. Ins.}, 24(6):891--896, 1989.

\bibitem{Latham1995}
Edited by~Rod~Latham.
\newblock {\em {High Voltage Vacuum Insulation}}.
\newblock Academic Press, London, 1995.

\bibitem{Williams1972}
D.~W. Williams and W.~T. Williams.
\newblock Effect of electrode surface finish on electrical breakdown in vacuum.
\newblock {\em J. Phys. D}, 5:1845--1854, 1972.

\bibitem{BastaniNejad2014}
M.~BastaniNejad, A.~A. Elmustafa, E.~Forman, J.~Clark, S.~Covert, J.~Grames,
  J.~Hansknecht, C.~Hernandez-Garcia, M.~Poelker, and R.~Suleiman.
\newblock Improving the performance of stainless-steel {DC} high voltage
  photoelectron gun cathode electrodes via gas conditioning with helium or
  krypton.
\newblock {\em Nucl. Instrum. Meth. A}, 762:135--141, 2014.

\bibitem{Sudarshan1988}
N.~C. Jaitly and T.~S. Sudarshan.
\newblock X-ray emission and prebreakdown currents in plain and dielectric
  bridged vacuum gaps under dc excitation.
\newblock {\em IEEE Transactions on Electrical Insulation}, 23(2):231--242, Apr
  1988.

\bibitem{HudsonThesis}
Eric~R. Hudson.
\newblock {\em Experiments on Cold Molecules Produced via Stark Deceleration}.
\newblock PhD thesis, University of Colorado, 2006.

\bibitem{NISTestar}
M.A.~Zucker M.J.~Berger, J.S.~Coursey and J.~Chang.
\newblock Stopping-power and range tables for electrons, protons, and helium
  ions.
\newblock Technical report, NIST, 2015.

\bibitem{Kramers1923}
H.~A. Kramers.
\newblock {XCIII}. on the theory of {X-ray} absorption and of the continuous
  {X-ray} spectrum.
\newblock {\em Philosophical Magazine Series 6}, 46(275):836--871, 1923.

\bibitem{NISTXray}
J.~H. Hubble and S.~M. Seltzer.
\newblock Tables of {X-ray} mass attenuation coefficients and mass
  energy-absorption coefficients from 1 {keV} to 20 {MeV} for elements ${Z}=1$
  to 92 and 48 additional substances of dosimetric interest.
\newblock Technical report, NIST, 1996.

\bibitem{Jackson}
J~D Jackson.
\newblock {\em Classical Electrodynamics}.
\newblock John Wiley \& Sons, third edition, 1999.

\bibitem{Note1}
For convenience, note that 1~nW of radiation absorbed by a body of 100~kg is
  equivalent to a dose rate of 36~nSv~hr$^{-1}$ (assuming a quality factor and
  tissue weighting factor of 1).

\bibitem{Baron2014}
The~ACME Collaboration, J.~Baron, W.~C. Campbell, D.~DeMille, J.~M. Doyle,
  G.~Gabrielse, Y.~V. Gurevich, P.~W. Hess, N.~R. Hutzler, E.~Kirilov,
  I.~Kozyryev, B.~R. O’Leary, C.~D. Panda, M.~F. Parsons, E.~S. Petrik,
  B.~Spaun, A.~C. Vutha, and A.~D. West.
\newblock Order of magnitude smaller limit on the electric dipole moment of the
  electron.
\newblock {\em Science}, 343(6168):269--272, 2014.

\bibitem{Wu2009}
Y.~Wu and FDS Team.
\newblock {CAD}-based interface programs for fusion neutron transport
  simulation.
\newblock {\em Fusion Engineering and Design}, 84(7--11):1987--1992, 2009.

\bibitem{Ludlum449}
Ludlum 44-9 pancake gm detector product manual.
\newblock Technical report, Ludlum Measurements Inc., 2014.

\bibitem{RadeyeManual}
{\em RadEye B20/B20-ER Multi-Purpose Survey Meter Manual}.

\bibitem{Gleichauf1951}
P.~H. Gleichauf.
\newblock Electrical breakdown over insulators in high vacuum.
\newblock {\em J. Appl. Phys.}, 22:766--771, 1951.

\bibitem{Miller1989}
H.~C. Miller.
\newblock Surface flashover of insulators.
\newblock {\em IEEE Trans. Electr. Insul.}, 24:765--786, 1989.

\bibitem{Shannon1965}
J.~P. Shannon, S.~F. Philp, and J.~G. Trump.
\newblock Insulation of high voltage across solid insulators in vacuum.
\newblock {\em J. Vac. Sci. Technol.}, 2:234--239, 1965.

\bibitem{Kuffel1972}
E.~Kuffel, S.~Grzybowski, and R.~B. Ugarte.
\newblock Flashover across polyethylene and tetrafluoroethylene surfaces in
  vacuum under direct, alternative and surge voltages of various waveshapes.
\newblock {\em J. Phys. D}, 5:575--579, 1972.

\bibitem{Grzybowski1983}
S.~Grzybowski and J.~E. Thompson.
\newblock Electric surface strength and surface deterioration of thermoplastic
  insulators in vacuum.
\newblock {\em IEEE Trans. Electr. Insul.}, EI-18:301--309, 1983.

\bibitem{Korzekwa1989}
R.~Korzekwa, F.~M. Lehr, H.~G. Krompholz, and M.~Kristiansen.
\newblock Inhibiting surface flashover for space conditions using magnetic
  fields.
\newblock {\em IEEE Trans. Plasma Sci.}, 17:612--615, 1989.

\bibitem{Hawley1968}
R.~Hawley.
\newblock Solid insulators in vacuum: A review (invited paper).
\newblock {\em Vacuum}, 18:383--390, 1968.

\bibitem{Faircloth2014}
D.~C. Faircloth.
\newblock {\em {Technological Aspects: High Voltage}}, pages 381--419.
\newblock CERN, 2014.

\bibitem{Bommakanti1990}
R.~G. Bommakanti and T.~S. Sudarshan.
\newblock Influence of mechanical grinding and polishing operations of brittle
  polycrystalline alumina on the pulsed surface flashover performance.
\newblock {\em J. Appl. Phys.}, 67:6991--6997, 1990.

\bibitem{Kato2007}
K.~Kato, Y.~Fukuoka, H.~Saitoh, M.~Sakaki, and H.~Okubo.
\newblock Effect of electrode surface roughness on breakdown conditioning under
  non-uniform electric field in vacuum.
\newblock {\em IEEE Trans Dielectr. Electr. Insul.}, 14:538--543, 2007.

\bibitem{Choulkov2005}
V.~V. Choulkov.
\newblock Effect of electrode surface roughness on electrical breakdown in high
  voltage apparatus.
\newblock {\em IEEE Trans. Dielectr. Electr. Insul.}, 12:98--103, 2005.

\bibitem{Hatfield1988}
L.~L. Hatfield.
\newblock A treatment which improves surface withstand voltage in vacuum.
\newblock {\em IEEE Trans. Electr. Insul.}, 23(1):57--61, 1988.

\bibitem{Dylla1988}
H.~F. Dylla.
\newblock Glow discharge techniques for conditioning high-vacuum systems.
\newblock {\em J. Vac. Sci. Technol. A}, 6:1276--1287, 1988.

\bibitem{Anjali1994}
S.~Anjali and S.~V. Gogawale.
\newblock Study of the parametric variation for argon glow discharge
  conditioning of {SS304}.
\newblock {\em Vacuum}, 45:41--44, 1994.

\bibitem{Yamamoto1989}
O.~Yamamoto, T.~Hara, T.~Nakae, and M.~Hayashi.
\newblock Effects of spark conditioning, insulator angle and length on surface
  flashover in vacuum.
\newblock {\em IEEE Trans. Electr. Insul.}, 24:991--994, 1989.

\bibitem{Hernandez-Garcia2009}
C.~Hernandez-Garcia, S.~V. Benson, G.~Biallas, D.~Bullard, P.~Evtushenko,
  K.~Jordan, M.~Klopf, D.~Sexton, C.~Tennant, R.~Walker, and G.~Williams.
\newblock {DC} high voltage conditioning of photoemission guns at {Jefferson
  Lab FEL}.
\newblock {\em AIP Conference Proceedings}, 1149:1071--1076, 2009.

\end{thebibliography}
\bibliographystyle{unsrt}

\end{document}